\documentclass[aps,prl,reprint,groupedaddress]{revtex4-1}

\usepackage{graphicx}
\usepackage{epstopdf}
\usepackage{dcolumn}
\usepackage{bm}
\usepackage{color}
\usepackage{amsfonts}
\usepackage{amsmath}  
\usepackage{xcolor}
\usepackage{soul}
\usepackage[colorlinks=true,citecolor=blue]{hyperref}
\hypersetup{colorlinks=true,citecolor=blue,linkcolor=blue,urlcolor=blue}

\begin{document}


\title{The structure of a perturbed magnetic reconnection electron diffusion region}




\author{G. Cozzani}
\email{giuliac@irfu.se}
\affiliation{Swedish Institute of Space Physics, Uppsala, Sweden}
\author{Yu.~V. Khotyaintsev}
\affiliation{Swedish Institute of Space Physics, Uppsala, Sweden}
\author{D.~B. Graham}
\affiliation{Swedish Institute of Space Physics, Uppsala, Sweden}
\author{J. Egedal}
\affiliation{Department of Physics, University of Wisconsin-Madison, Madison, Wisconsin, USA}
\author{M. Andr\'e}
\affiliation{Swedish Institute of Space Physics, Uppsala, Sweden}
\author{A. Vaivads}
\affiliation{Department of Space and Plasma Physics, School of Electrical Engineering and Computer Science, KTH Royal Institute of Technology, Stockholm, Sweden}
\author{A. Alexandrova}
\affiliation{Laboratoire de Physique des Plasmas, CNRS, Sorbonne Université, Université Paris-Saclay, Observatoire de Paris, \'Ecole Polytechnique Institut Polytechnique de Paris, Palaiseau, France}
\author{O. Le Contel}
\affiliation{Laboratoire de Physique des Plasmas, CNRS, Sorbonne Université, Université Paris-Saclay, Observatoire de Paris, \'Ecole Polytechnique Institut Polytechnique de Paris, Palaiseau, France}
\author{R. Nakamura}
\affiliation{Space Research Institute, Austrian Academy of Sciences, Graz, Austria}
\author{S. A. Fuselier}
\affiliation{Southwest Research Institute, San Antonio, Texas, USA}
\author{C. T. Russell}
\affiliation{University of California, Los Angeles, California, USA}
\author{J. L. Burch}
\affiliation{Southwest Research Institute, San Antonio, Texas, USA}
\date{\today}

\begin{abstract}
We report in situ observations of an electron diffusion region (EDR) and adjacent separatrix region. We observe significant magnetic field oscillations near the lower hybrid frequency which propagate perpendicularly to the reconnection plane. We also find that the strong electron-scale gradients close to the EDR exhibit significant oscillations at a similar frequency. Such oscillations are not expected for a crossing of a steady 2D EDR, and can be explained by a complex motion of the reconnection plane induced by current sheet kinking propagating in the out-of-reconnection-plane direction. Thus all three spatial dimensions have to be taken into account to explain the observed perturbed EDR crossing.
\end{abstract}


\maketitle


Magnetic reconnection is a fundamental plasma process that yields to the topological reconfiguration of the magnetic field and the concurrent energization and acceleration of plasma species \citep{Vasyliunas1975}. Reconnection is found in a variety of environments in space and astrophysical plasmas \citep{ZweibelYamada2009} and dedicated laboratory experiments \citep{Yamada1997,Forest2015}. 
A crucial constituent of the collisionless reconnection process is the electron diffusion region (EDR), where the demagnetization of both ions and electrons enables the magnetic field topology change. As a result, the processes that take place in the EDR affect the system up to its global MHD scales.
Despite their central role, these processes are still largely unknown. In particular, the contribution of plasma waves and instabilities to the EDR dynamics as well as to the overall reconnection process remain unclear \citep{Fujimoto2011,Khotyaintsev2019}. Waves and instabilities operating in the center of the current sheet could affect the two-dimensional, steady and laminar reconnection picture. For guide-field reconnection, in particular, the role of streaming instabilities leading to turbulence development at the reconnection site has been discussed in simulation studies \citep{Che2017, Drake2003} and electrostatic turbulence promoting electron heating is observed at a magnetopause EDR \citep{Khotyaintsev2020}.

Among the instabilities that can develop in current layers, the lower hybrid drift instability (LHDI) has been extensively studied since it can potentially provide anomalous resistivity sustaining the reconnection electric field \citep{Huba1977}. Some early observational work supported this idea \citep{Cattell1995}. 
However, spacecraft observations at the magnetopause \citep{Bale2002,Vaivads2004,Graham2017} and magnetotail \citep{Eastwood2009,Zhou2009}  suggest that electrostatic LHDI modes could not supply the necessary resistivity, consistent with the fact that these modes develop at the edges of the current sheet but are stabilized in the center \citep{Davidson1977}. On the other hand, eigen-mode analysis and kinetic simulations of ion-scale Harris current sheets \citep{Daughton2003,Yoon2002} suggest that electromagnetic LHDI modes can penetrate in to the center of the current layer. Such modes are characterised by lower growth rates and longer wavelength compared to the electrostatic modes. Electromagnetic fluctuations in the lower-hybrid frequency range were observed within a reconnecting current sheet in the MRX laboratory experiment \citep{Ji2004} but in situ observations of electromagnetic LHDI modes within the EDR are still lacking.

Indeed, before the launch of the Magnetospheric Multiscale (MMS) mission \citep{Burch2016mms}, observational evidence of these instabilities occurring at the EDR were prevented by the lower resolution of the available particle measurements and by the limited knowledge of the EDR and related electron-scale processes. Electrostatic lower hybrid drift waves (LHDW) in the EDR have been investigated only recently \citep{Chen2020}.

In this Letter, we report MMS observations of a magnetotail electron diffusion region and adjacent separatrix region characterised by unexpected electric field, electron velocity and magnetic field oscillations. We compare 2D fully kinetic simulations and four-spacecraft observations to investigate the mechanism responsible for the observed oscillations.

MMS encountered an EDR on August 10, 2018 at 12:18:33 UTC when it was located in the Earth's magnetotail at $[-15.2, \ 4.6, \ 3.1]_{GSM} \ R_E$ (in Geocentric Solar Magnetospheric system). The indicative signatures of an EDR \citep{Burch2016,Webster2018,Torbert2018} -- including super-Alv\'enic electron jets, enhanced electron agyrotropy, intense energy conversion and crescent-shaped electron distribution functions -- are observed \citep{Zhou2019}. During this event, MMS stays mostly in the plasma sheet ($B \sim 7 \ nT$ and $n \sim 0.17 \ cm^{-3}$). A weak guide field $B_g \sim 2 \ nT \sim 0.13 B_{inflow}$ is present ($B_{inflow}$ is the inflow magnetic field computed in the interval 12:21:20-12:21:40 \citep{Zhou2019}). 
The mean inter-spacecraft separation $\sim 20 \ km$ is comparable to the electron inertial length $d_e \sim 13 \ km$. As a first step, we determine the appropriate LMN coordinate system and establish the MMS trajectory relative to the EDR by adopting methods reported in Refs.\citep{Egedal2019,Shuster2017}. For this we use a 2D-3V kinetic PIC simulation performed with the VPIC code \citep{Bowers2008} which mimics the MMS event in terms of guide field (simulation run featuring upstream $\beta_{e, \infty} = 0.09$ and $B_g = 0.1$ \citep{Le2013}). The realistic ion-to-electron mass ratio $m_i/m_e = 1836$ allows us to establish a one-to-one correspondence between the dimensionless units of the simulation and the physical units of MMS data.  

\begin{figure}
    \centering
    \includegraphics[width=1\columnwidth]{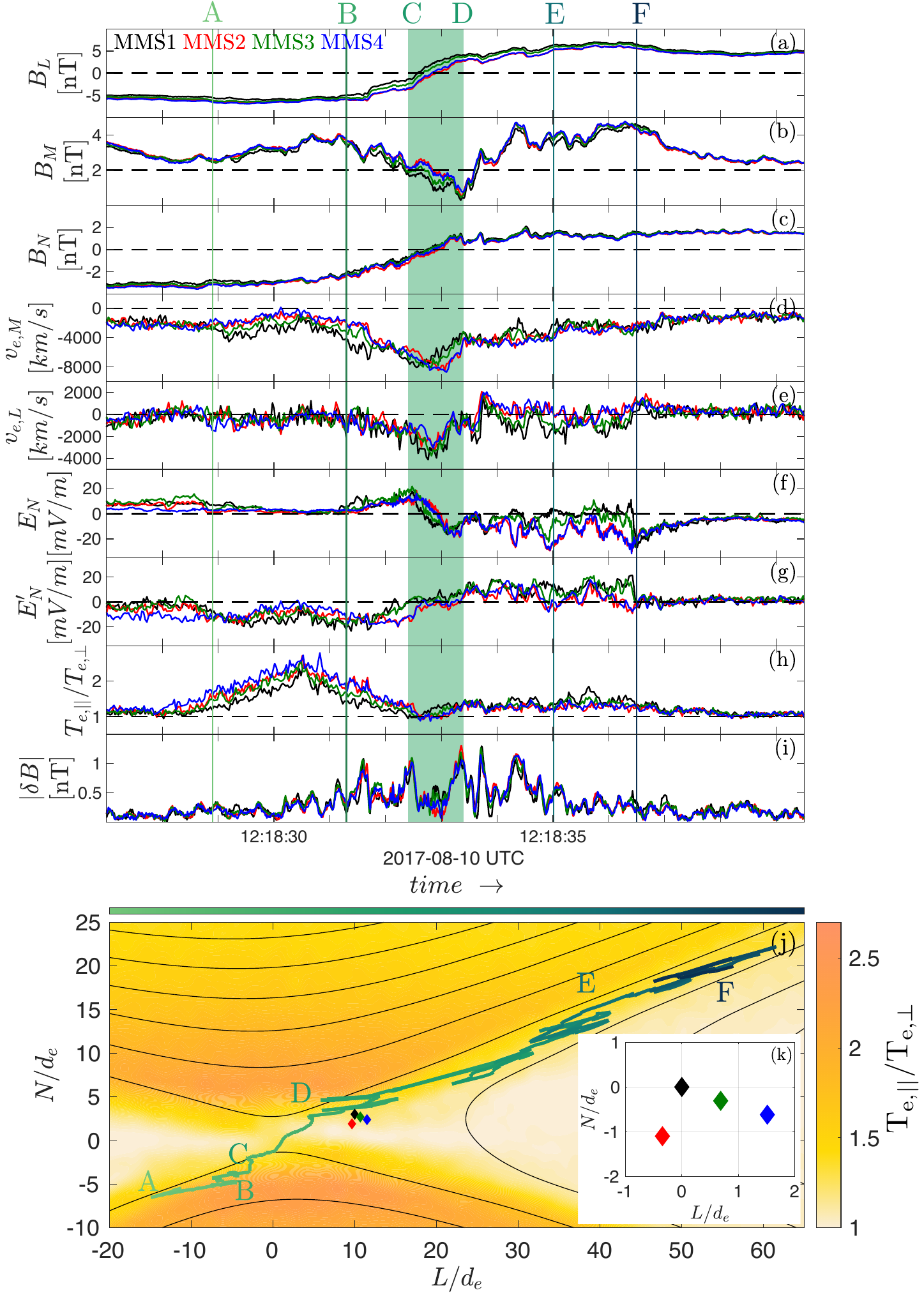}
    \caption{Top: Four spacecraft (a) $B_L$, (b) $B_M$ and (c) $B_N$ measured by FGM \citep{Russell2016}; (d) $v_{e,M}$ and (e) $v_{e,L}$ from FPI \citep{Pollock2016}; (f) $E_{N}$ from EDP \citep{Ergun2016,Lindqvist2016}; (g) N component of $\mathbf{E}' = \mathbf{E} + \mathbf{v}_e \times \mathbf{B}$; (h) $T_{e,||}/T_{e,\perp}$; (i) magnitude of magnetic field fluctuations (computed filtering FGM data with $0.5 \ Hz < f < 64 \ Hz$). The green shaded interval indicates the inner EDR. Bottom: (j) 2D PIC simulation data of $T_{e,||}/T_{e,\perp}$ with the reconstructed MMS trajectory crossing the EDR. The magnetic flux contour lines are superposed; (k) spacecraft position relative to MMS1 in the LN plane. \label{fig1}}
\end{figure}

Fig.\ref{fig1} shows an overview of the EDR crossing. All the quantities are shown in the LMN coordinate system ($\mathbf{L} = [0.96, \   -0.15, \ -0.22]$, $\mathbf{M} = [0.17, \ 0.98, \    0.03]$, $\mathbf{N} = [0.22, \ -0.07, \ 0.97]$ in GSM, obtained via an optimisation approach aided by simulation data \citep{Egedal2019}). The MMS trajectory relative to the EDR is shown in Fig.\ref{fig1}(j). The trajectory is reconstructed in interval A--F (12:18:28.9-12:18:36.5) of Fig.\ref{fig1}. The part of trajectory corresponding to interval 12:18:28.9-12:18:34.8 is reconstructed by adapting the method of Ref.\citep{Shuster2017} to include the electron velocity $v_{e,M}$ and the electron temperature anisotropy. For the part of the trajectory corresponding to interval 12:18:34.8-12:18:36.5, we use the method of Ref.\citep{Egedal2019} (including $E_N$ and $B_L$) which allows us to reproduce the observed electric field oscillations.  

MMS is initially located south and tailward of the reconnection site, corresponding to $B_L < 0$ (Fig.\ref{fig1}(a)), $B_N < 0$ (Fig.\ref{fig1}(c)) and $V_{i,L} < 0$ (not shown). Then, MMS crosses the diffusion region diagonally so that $B_L$ and $B_N$ change from negative to positive. MMS samples mainly the positive lobes of the Hall quadrupolar field ($B_M > B_g$, Fig.\ref{fig1}(b)). Figure \ref{fig1}(h) shows the electron temperature anisotropy $T_{e,||}/T_{e,\perp}$, where parallel and perpendicular refer to the local magnetic field direction. The $T_{e,||}/T_{e,\perp}$ peak observed at 12:18:30.5 indicates that MMS performed a brief excursion into the inflow region, where $T_{e,||}/T_{e,\perp}$ is expected to increase \citep{Egedal2008, Egedal2013}, before approaching the inner EDR (interval C--D) \citep{Karimabadi2007}. Interestingly, during the current sheet crossing (interval B--E), MMS observes significant magnetic field oscillations $\delta B$ (Fig.\ref{fig1}(i)) reaching $\sim 20 \%$ of the upstream magnetic field in the plasma sheet ($\sim 7 \ nT$). Applying the timing method \citep{Harvey1998} on the sharp $B_L$ variation in interval  12:18:32.0 - 12:18:33.3 we estimate the current layer width to be $d_{cs} \sim 2 \ d_e$, in agreement with Ref.\citep{Zhou2019}. This implies that MMS crossed an electron scale current sheet.

While the typical signatures of an EDR encounter are observed overall, the multi-spacecraft analysis of electric and velocity fields along the spacecraft trajectory allows us to identify signatures which are distinctive of this event.  
Figure \ref{fig1}(f) shows the normal component of the electric field, $E_N$, exhibiting a bipolar behavior (positive on the -N side and negative on the +N side of the neutral line) consistent with Hall dynamics. While the different spacecraft see similar $E_N$ in the interval A--D (12:18:28.9 - 12:18:33.4),  significant differences between the spacecraft are observed in interval D--F (12:18:33.4 - 12:18:36.5). Indeed, while MMS2 and MMS4 observe $E_N < 0$, MMS1 measures $E_N \sim 0$ and even $E_N > 0$. The largest difference is observed between spacecraft with the largest separation in the N direction (MMS1 and MMS2, Fig.\ref{fig1}(k)) while spacecraft which are close to each other in the N direction and separated both in L direction observe nearly identical signals. MMS2-MMS4 is the spacecraft pair with the largest separation in the M direction ($1.1 \ d_e$, not shown) but the observations from the two spacecraft are nearly identical. We conclude that the observed differences at the scale of the tetrahedron are related to different positions primarily in the N direction.

The difference between $E_N$ measured at MMS1 and MMS2 (which are only $1.1 \ d_e$ apart along the N direction) reaches a maximum value of $\sim 30 \ mV/m$ (e.g. at 12:18:35.01). This indicates the presence of strong gradients at the electron scales. 
Analogously to the differences in $E_N$, also significant differences are observed in $v_{e,L}$ (Fig.\ref{fig1}(e)), reaching 2000 km/s, and in the parameter $\mathbf{E}' = \mathbf{E} + \mathbf{v}_e \times \mathbf{B}$ (Fig.\ref{fig1}(g)) which quantifies the demagnetization of the electrons. $E'_N \neq 0$ for the majority of interval A--F, indicating that the electrons are not frozen-in to the magnetic field. These differences further confirm the presence of strong gradients on spatial scales $\sim d_e$.

Hence, during this EDR encounter we identify strong electron scale gradients and electron demagnetization. However, the most intriguing feature of this EDR crossing is the presence of large fluctuations in $E_N$, $v_{e,L}$ along the separatrix (region D--F) and of $\delta \mathbf{B}$ in the center of the current sheet (interval B--E). Such oscillations are not expected for a smooth crossing of a laminar EDR, and their presence indicates that the EDR crossing is perturbed by some process. We investigate these oscillations in detail in order to identify this process. 

\begin{figure}
    \centering
    \includegraphics[width=1\columnwidth]{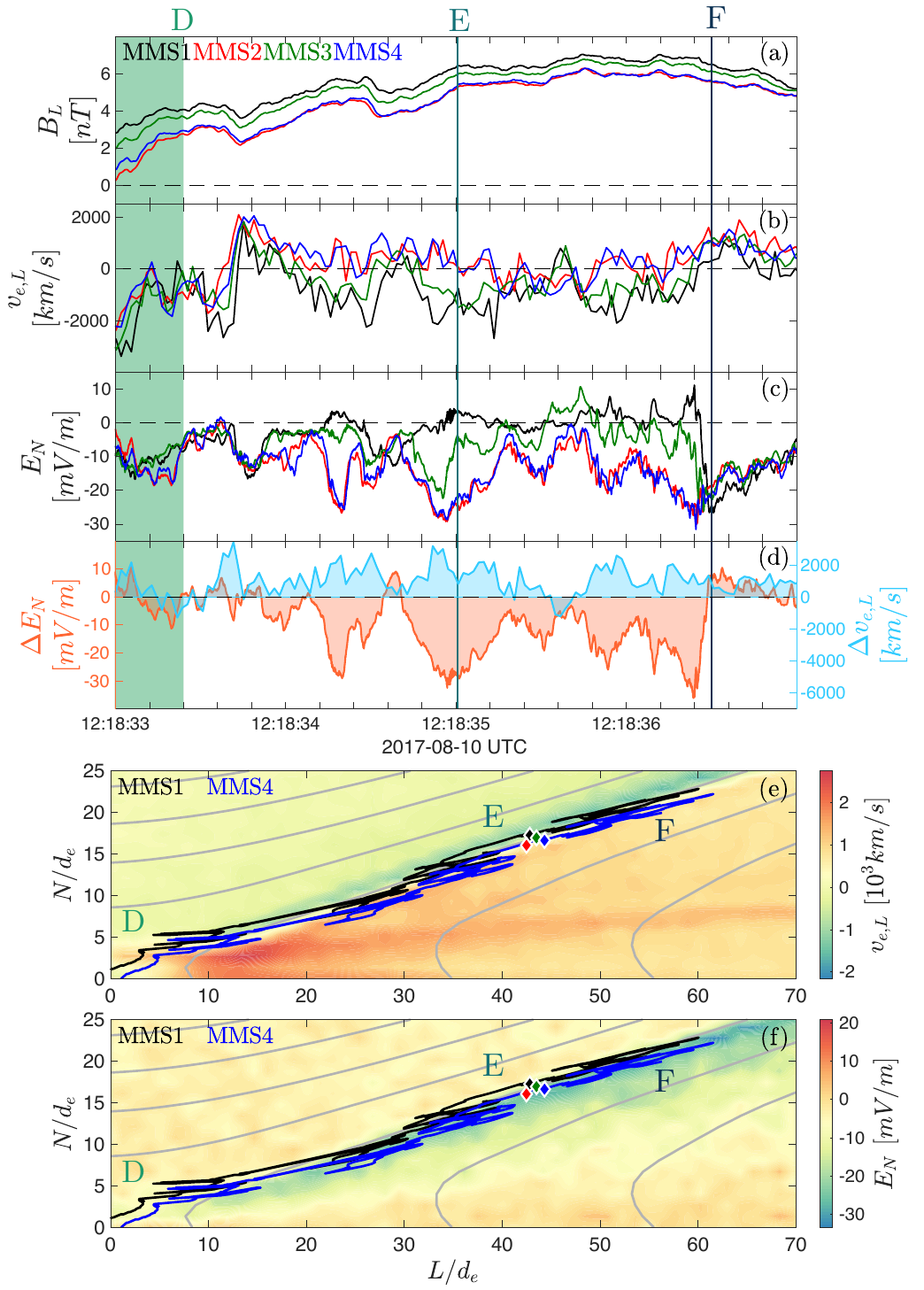}
    \caption{Four spacecraft MMS observations: (a) $B_{L}$; (b) $v_{e,L}$; (c) $E_{N}$; (d) $\Delta E_N = E_{N,MMS2} - E_{N,MMS1}$ and $\Delta v_{e,L} = v_{e,L,MMS2} - v_{e,L,MMS1}$. 
PIC simulation: (e) contour plot of $v_{e,L}$; (f) contour plot of $E_{N}$. The black and blues lines represent the MMS1 and MMS4 trajectories. \label{fig2}}
\end{figure}

Figure \ref{fig2} focuses on the separatrix region characterised by the strong gradients. Both $v_{e,L}$ and $E_N$ (Fig.\ref{fig2}(b)--(c)) show very different profiles at each of the spacecraft. Notably, MMS2 and MMS4 observe a strongly fluctuating and mostly negative $E_N$ while the $E_N$ is mostly positive for MMS1 and the fluctuations are not as prominent. Indeed, the observed difference between $E_{N}$ measured by MMS1 and MMS2 ($\Delta E_N = E_{N,MMS2} - E_{N,MMS1}$, Fig.\ref{fig2}(d)) and analogously between $v_{e,L}$ measured by MMS1 and MMS2 ($\Delta v_{e,L} = v_{e,L,MMS2} - v_{e,L,MMS1}$) show large variations.
Such large variations in the observed gradients can be either caused by kinking of the current sheet as a whole or by temporal variations of the gradients at electron scales, or by a combination of the two.   

Figures \ref{fig2}(e)--(f) show 2D PIC simulation data of $E_N$ and $v_{e,L}$ in the LN plane. The location of  MMS corresponding to the E-labeled line in Fig.\ref{fig2}(a)--(d) is shown in the LN plane. The simulation data (Fig. \ref{fig2}(e)--(f))  exhibit large differences in $E_N$ and $v_{e,L}$ at the different spacecraft locations, thus electron scale gradients as the ones identified in the in situ observations are also present in the simulation data.
However, considering the laminar character of the simulation data, if one were to consider a smooth MMS trajectory across a steady-state 2D reconnection plane (see e.g. \citep{Torbert2018,Egedal2019}), one would expect the difference between $E_N$ and $v_{e,L}$ observed at different spacecraft to be rather constant and the related gradients to be uniform along the separatrix.
This is in striking contrast with the large variations in the gradients observed by MMS. The 2D simulation can be matched to the in situ data only if we use a rather complex trajectory, as shown in Fig. \ref{fig2}(e)--(f)).  This trajectory is overall tangential to the separatrix, yet it exhibits several back-and-forth motions which are necessary to reproduce the oscillating $\Delta E_N$ and $\Delta v_{e,L}$ observed in situ.
 
\begin{figure}
    \centering
    \includegraphics[width=1\columnwidth]{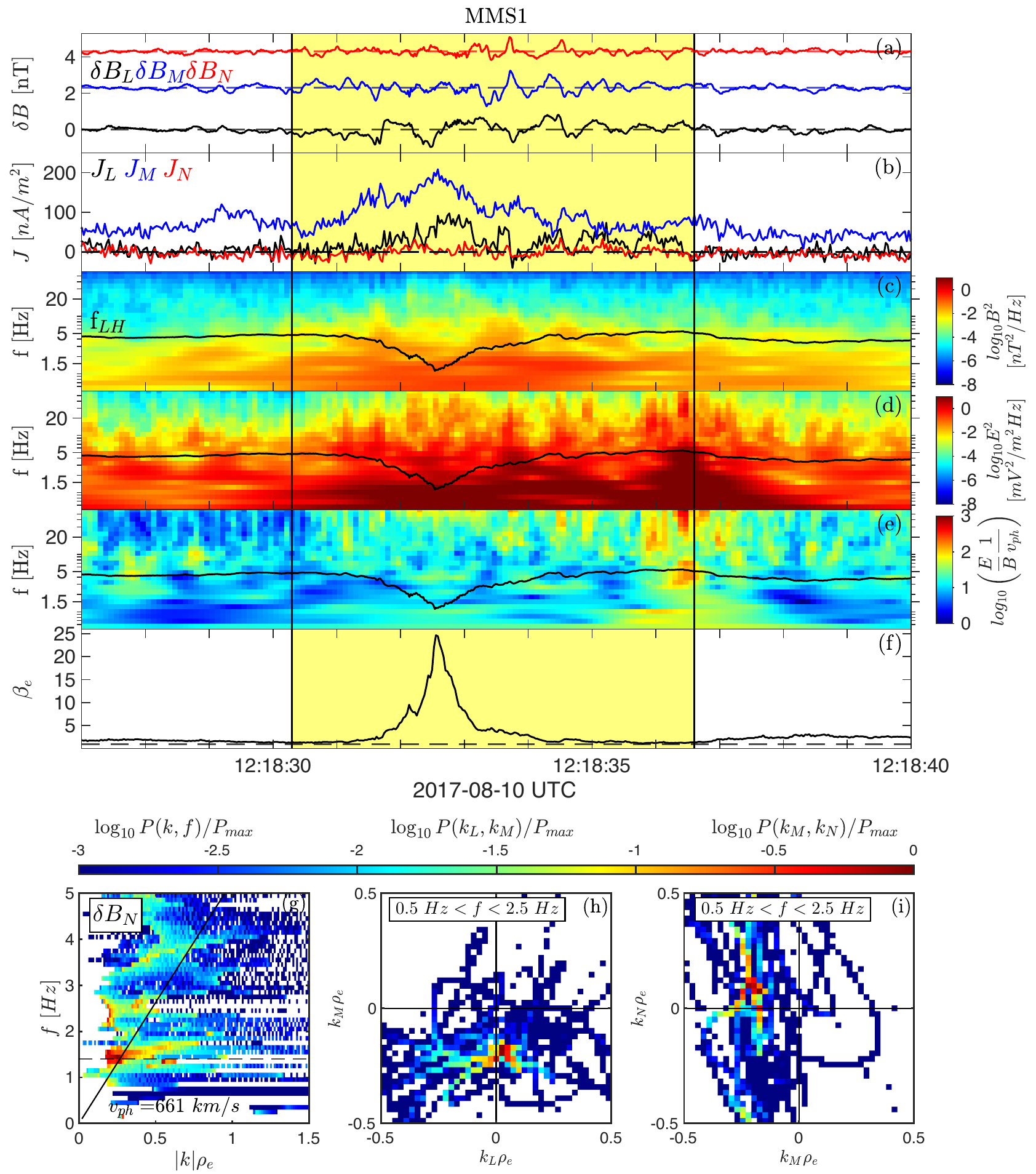}
    \caption{Top: (a) Three components of $\delta \mathbf{B}$. Offsets of 2.3 nT and 4.3 nT are added to $\delta B_M$ and $\delta B_N$ respectively; (b) three components of $J$ calculated from particle moments; (c) Spectrum of $\mathbf{B}$ wave power; (d) spectrum of the $\mathbf{E}$ wave power; (e) spectrum of $\log_{10} \left( \frac{E}{B} \frac{1}{v_{ph}} \right)$; (f) $\beta_e$. The black line indicates $f_{LH}$. Bottom: Normalized power of magnetic field fluctuations $\delta B_N$ versus (g) $|k| \rho_e$ and frequency; (h)  $k_{L} \rho_e$ and $k_M \rho_e$ ($0.5 \ Hz < f < 2.5 \ Hz$); (i) $k_{M} \rho_e$ and $|k_N| \rho_e$ ($0.5 \ Hz < f < 2.5 \ Hz$);. The dashed line in panel (g) corresponds to $f = 1.4 \ Hz$.\label{fig3}}
\end{figure}

In order to identify the process responsible for the complex EDR crossing, we analyze the observed fluctuations of magnetic field $\delta \mathbf{B}$ (see Fig.\ref{fig1}(i)).
 Fig.\ref{fig3}(a) shows that the $\delta \mathbf{B}$ fluctuations, with similar amplitude in all three components,  are present in the center of the current sheet, where the current density peaks (Fig.\ref{fig3}(b), yellow shaded interval 12:18:30.3 - 12:18:36.5). Figure \ref{fig3}(c) and \ref{fig3}(d) show the wavelet power spectra of the electric and magnetic fields observed by MMS1. Both the magnetic and electric powers clearly drop for frequencies $f > f_{LH}$ ($f_{LH} \approx \sqrt{f_{ci} f_{c,e}}$ is the lower hybrid frequency) and in the inner EDR the waves have $f \sim f_{LH}$.
The parameter $\frac{E}{B} \frac{1}{v_{ph}}$ (Fig.\ref{fig3}(e)), where $v_{ph}$ is the phase speed of the observed waves (see Fig.\ref{fig3}(g)), is used to quantify the electrostatic and electromagnetic component of the waves. Theoretically, the parameter $\frac{E}{B} \frac{1}{v_{ph}} \to \infty$ for purely electrostatic waves. Averaging this parameter in the yellow shaded interval of Fig.\ref{fig3} and in the frequency range $1 \ Hz < f < 5 \ Hz$, we obtain a mean value of $\frac{E}{B} \frac{1}{v_{ph}}\sim 15$ which is much smaller than the typical value of $\frac{E}{B} \frac{1}{v_{ph}}$ in the quasi electrostatic case. For example, $\frac{E}{B} \frac{1}{v_{ph}} \sim 400$ (for $0.3 < f/f_{LH} < 0.8$) for the quasi-electrostatic fluctuations reported in Ref.\citep{Graham2019}.  Thus, the fluctuations in the center of the reconnecting current sheet are characterised by a significant electromagnetic component.

To better characterize these fluctuations, we compute the dispersion relation from the phase differences of $\delta B_N$ between spacecraft pairs, using multi-spacecraft interferometry \citep{Graham2016, Graham2019}. Figure \ref{fig3}(g) shows that the normalized power $P(f,k)/P_{max}$ peaks at $f \sim 1.4 \ Hz$ (black dashed line) which is close to $f_{LH}$ at the current sheet center (12:18:32.8). The wave number at the $P(f,k)/P_{max}$ peak is $k \rho_e \sim 0.3$ ($\rho_e \sim 24 \ km$ is the electron gyroradius) which corresponds to phase speed $v_{ph} = 660 \ km/s$ and wavelength $\lambda \sim 500 \ km$. 
Figure \ref{fig3}(h)--(i) shows that the wave vector $\mathbf{k}$ is directed mainly along the M direction, i.e. it is anti-aligned with the direction of the current and perpendicular to the reconnection plane LN. The average direction of propagation of the fluctuations is $\mathbf{\hat{k}} = [0.12, \  - 0.92, \   0.38]$ in LMN coordinates and it is mainly perpendicular to the magnetic field direction ($\theta_k  = \arccos {\frac  {{\mathbf  {k}}\cdot {\mathbf  {B}}}{|{\mathbf  k}||{\mathbf  B}|}} \sim 70^{\circ}$, not shown).  Similar results are obtained if a different component of $\delta \mathbf{B}$ is considered for the analysis. 
These signatures are consistent with lower hybrid drift fluctuations propagating in the out-of-reconnection-plane direction.

\begin{figure}
    \centering
    \includegraphics[width=1\columnwidth]{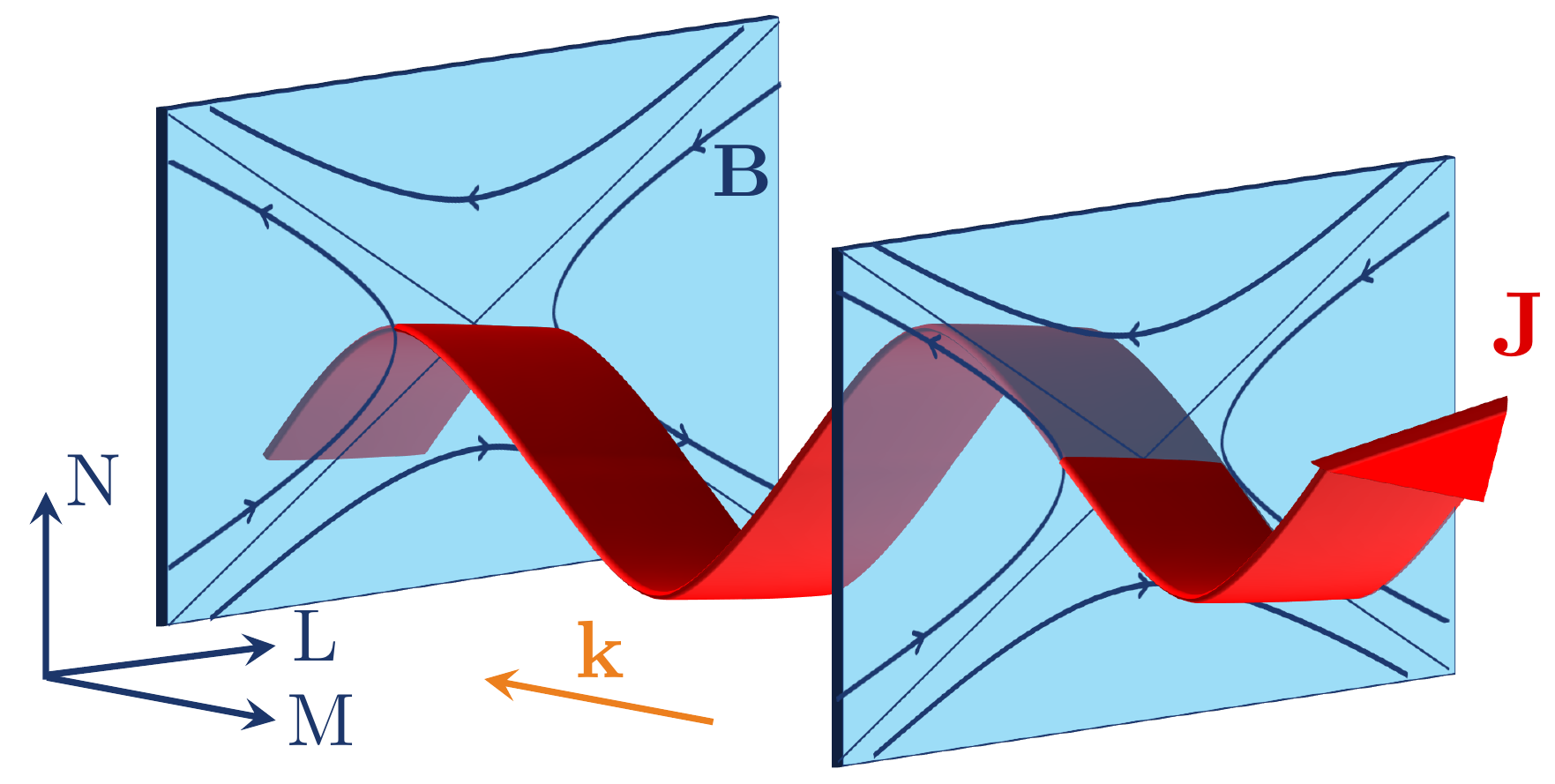}
    \caption{Schematic representation of the kinking of the electron scale current sheet propagating in the out-of-reconnection-plane direction (not to scale). \label{fig4}}
\end{figure}

The $\delta \mathbf{B}$ fluctuations in the current sheet center and the electric and velocity field fluctuations at the separatrix have similar time scales which are comparable to the lower hybrid frequency (Fig. \ref{fig3}(c)--(d) and \ref{fig2}(d)). This similarity suggests that they are related to each other. 
As shown in Fig.\ref{fig2}, we can match the observed oscillating $\Delta E_N$ and $\Delta v_{e,L}$ to the steady-state 2D reconnection structure if we employ a complex motion of the 2D reconnection plane. Both such complex motion and the $\delta \mathbf{B}$ fluctuations in the center of the current sheet can be produced by kinking of the current sheet propagating in the out-of-reconnection-plane direction (see a qualitative representation in Fig.\ref{fig4}). On the other hand, given the electron-scale inter-spacecraft separation which does not allow the sampling of the larger scales, we cannot establish whether the oscillations shown in Fig.\ref{fig2}(b)--(d) are indeed produced exclusively by the rigid motion of the reconnection plane, or if a more complex behavior including time evolution is present.

The fluctuations observed during the EDR crossing are related to one of the various drift instabilities that are eigen-oscillations resulting in current sheet kinking \cite{Daughton2003,Yoon2002}. Several modes that have been considered as distinguished in the past actually belong to the same class of instabilities ranging from the electrostatic lower hybrid drift instability LHDI (fast growing, short-wavelength mode with  $k \rho_e \sim 1$) localized at the edges of the current sheet \citep{Davidson1977} to the electromagnetic, longer-wavelength  modes with $k \sqrt{\rho_i \rho_e} \sim 1$ located close to the current sheet center which arise in later phases of the instability \cite{Daughton2003,Yoon2002, Shinohara2001, Suzuki2002, Scholer2003}.  In the event reported here, MMS observed electromagnetic fluctuations with $k \rho_e \sim 0.3$ (which is somewhat smaller than the typical $k \rho_e \sim 0.5 - 1$ observed for LHDW at the magnetopause \citep{Khotyaintsev2016, Graham2017}) and $k \sqrt{\rho_i \rho_e} \sim 2.7$  located within the EDR ($\rho_e$ and $\sqrt{\rho_i \rho_e}$ are averaged over the yellow shaded interval of Fig.\ref{fig3}). These fluctuations are rather similar to the electromagnetic current sheet modes described in Ref.\citep{Daughton2003, Yoon2002}. 
Electromagnetic fluctuations in a reconnecting current sheet have been observed at the magnetic reconnection experiment (MRX) \citep{Ji2004}, and it was suggested that the fluctuations were generated by the Modified Two Stream Instability (MTSI) \citep{McBride1972,Hsia1979} which can occur at higher $\beta_e$ observed in the current sheet center (see Fig.\ref{fig3}(e)). 

Nonetheless, the comparison between our observations and the analytical/simulation studies \citep{Daughton2003,Yoon2002} or laboratory/spacecraft observations \citep{Ji2004, Asano2003} focusing on current sheet instabilities is constrained by the fact that the current sheet thickness in these studies is $d_{cs} \sim \ d_i$ while our event presents a very thin current sheet $d_{cs} \sim 2 \ d_e = 0.05 \ d_i$. Also, the plasma considered in previous studies is usually homogeneous \citep{Wu1983} or reconnection is not present \citep{Daughton2003,Yoon2002} or it is asymmetric \citep{Roytershteyn2012}. Independently of the specific instability operating in the current sheet, when the direction of propagation is perpendicular to the reconnection plane the out-of-plane direction cannot be treated as an invariant axis of the system. Thus, a 3D description is required to understand the dynamics of the process.

In conclusion, we report MMS observations of a perturbed EDR crossing. We observe oscillations of the electron-scale gradients at the separatrix and magnetic field fluctuations in the center of the current sheet. These features are not expected for a simple crossing of a steady-state 2D EDR. We find an overall good agreement between the observations and 2D PIC simulations of reconnection, but we can only match the observed oscillations to the 2D model if we consider a complex motion of the spacecraft in the fixed 2D reconnection plane. We attribute such complex motion to a kinking of the current sheet which is propagating in the out-of-reconnection-plane direction. Despite the overall quasi-2D geometry of the event, these results suggest that we need to take into account the three-dimensionality of the system to fully understand the observed EDR crossing. Further in situ data analysis and three-dimensional kinetic simulations enabling the out-of-plane dynamics are needed to establish the role of current sheet instabilities in affecting the EDR structure.

\begin{acknowledgments}
 We thank the entire MMS team and instruments PIs for data access and support. MMS data are available at the \href{https://lasp.colorado.edu/mms/sdc/public}{MMS data center}. We gratefully thank W. Daughton for running the simulations. This work was supported by the Swedish Research Council, Grants No. 2016-05507 and 2018-03569, and the Swedish National Space Agency, Grants No. 128/17 and 144/18. 
 
 \vspace{0.5mm}
 
 G.C. dedicates this work to the memory of Federico Tonielli.
\end{acknowledgments}

\end{document}